\documentclass[useAMS]{mn2e}
\usepackage{psfig}

\usepackage{times}
\usepackage{amssymb,amsmath}

\def\lsim{ \lower .75ex\hbox{$\sim$} \llap{\raise .27ex \hbox{$<$}} }
\def\gsim{ \lower .75ex \hbox{$\sim$} \llap{\raise .27ex \hbox{$>$}} }

\title[Magnetic fields of BL Lac jets] 
{On the magnetization of BL Lac  jets}

\author[Tavecchio \& Ghisellini]
{F. Tavecchio$^1$\thanks{E--mail:s fabrizio.tavecchio@brera.inaf.it} and
G. Ghisellini$^1$\\
$^1$INAF -- Osservatorio Astronomico di Brera, via E. Bianchi 46, I--23807
Merate, Italy\\
}

\begin{document}



\maketitle

\begin{abstract} 
The current paradigm foresees that relativistic jets are launched as magnetically dominated flows, 
whose magnetic power is progressively converted to kinetic power of 
of the matter of the jet, until equipartition is reached. 
Therefore, at the end of the acceleration phase, the jet should still carry a substantial fraction 
($\approx$ half) of its power in the form of a Poynting flux. 
It has been also argued that, in these conditions, the best candidate particle acceleration mechanism is 
 efficient reconnection of  magnetic field lines, for which it is predicted that  
magnetic field and accelerated relativistic electron energy densities are in equipartition.
Through the modeling of the jet non--thermal emission, we explore if equipartition 
is indeed possible in BL Lac objects, i.e. low--power blazars with weak or absent broad emission lines. 
We find that one--zone models (for which only one region is involved in the production of the radiation we observe) 
the particle energy density is largely dominating (by 1--2 orders of magnitude) over the magnetic one.
As a consequence, the jet kinetic power largely exceeds the magnetic power. 
Instead, if the jet is structured (i.e. made by a fast spine surrounded by a slower layer), the amplification of the IC emission due to the radiative interplay between the two components allows us to reproduce the emission in equipartition conditions.
\end{abstract}


\section{Introduction}

The current picture describing launching and acceleration processes of relativistic jets in 
active galactic nuclei (AGN) attributes a key role to magnetic fields, by means of which the 
energy stored in a rapidly spinning (Kerr) supermassive black hole (BH) can be extracted and 
channeled into a Poynting flux (Blandford \& Znajek 1977, Tchekhovskoy et al. 2009, 2011). 
The jet power, originally carried by an almost pure electromagnetic beam 
(with magnetization parameter $\sigma\equiv P_B/P_{\rm kin}\gg 1$), is progressively used to 
accelerate matter, until a substantial equipartition between the magnetic and the kinetic 
energy fluxes ($\sigma \approx 1$) is established (e.g., Komissarov et al. 2007, 2009, 
Tchekhovskoy et al. 2009, Vlahakis 2015). 
Dissipation of part of the kinetic (through shocks) and/or magnetic (through reconnection) 
power leads to the acceleration of particles up to ultra--relativistic energies, producing the non--thermal 
emission we observe from the jets of {\it blazars}, i.e. radio--loud AGN with a jet closely pointing 
toward the Earth (Urry \& Padovani 1995). 
This consistent paradigm has recently received strong support by extensive general relativistic 
MHD simulations (e.g. Tchekhovskoy et al. 2009, 2011), proving that the power actually extracted 
from the BH--accretion disk system is larger than the sole accreting power $\dot{M}c^2$, as indeed 
expected if the system can access the BH rotation energy, i.e. if the Blandford \& Znajek (1977) mechanism is at work. Observationally, this result is supported by studies comparing the jet and the accretion power, 
which show that the jet indeed carries a power larger than that associated to the accreted matter, 
estimated through the disk radiation (Ghisellini et al. 2010, 2014). Simulations demonstrate that 
a key role in determining such an efficient energy extraction is played by dynamically important 
magnetic fields built up in the inner regions of the accretion flow. 
In fact, the magnetic flux close to the BH horizon is so large 
that  accretion likely occurs through a magnetically arrested accretion (MAD) flow 
(Narayan et al. 2003, Tchekhovskoy et al. 2011, McKinney et al. 2012).

An important point of the scheme sketched above is that -- absent substantial dissipation -- the 
magnetic energy flux of the jet at large scale should still represent a large fraction 
($\approx 0.5$) of the total jet power. 
Ideally the magnetic flux carried by the jet should be comparable to that supported by inner accretion 
disk and regulating the energy extraction from the BH. 
From the observational point of view, the situation is however not clear.  
Zamaninasab et al. (2014) showed that the jet magnetic flux at parsec 
scale -- derived through the observed frequency-dependent core--shift in the radio band  
(Lobanov et al. 1998) -- correlates with the power of the corresponding accretion 
flow and has an absolute value comparable to that predicted by a MAD. 
An alternative and reliable way to estimate the magnetic fields associated to the emitting regions 
of the innermost ($\lesssim 1$ pc) part of jets is through the modeling of the relativistically 
beamed non--thermal continuum of blazars (e.g. Ghisellini et al. 1998, 2010, Tavecchio et al. 1998, 2010). 
Nalewajko, Sikora \& Begelman (2014) noted that the  large magnetic field flux derived through the 
core--shift effects by Zamaninasab et al. (2014) seems to exceed the values suggested by the 
typical blazar spectral energy distribution (SED). The great majority of the sources of the sample used by Zamaninasab et al. (2014) are flat spectrum radio quasars (FSRQ), in which the high--energy bump of the SED, thought to derive from the inverse Compton scattering by relativistic electrons in the jet, largely dominates over the low energy synchrotron bump, thus pointing to relatively small magnetic fields (though this fact does not directly implies {\bf sub-equipartition} magnetic fields, since the ratio of the IC and synchrotron luminosity -- related to the ratio of the radiation and magnetic energy densities --  does not directly involve the electron energy density).

Another line of research supporting the important role of magnetic fields in jets is that related to particle acceleration. Indeed, it is becoming clear that the widely assumed diffusive shock acceleration process 
(e.g. Heavens \& Drury 1988) is quite inefficient in accelerating highly relativistic 
particles with energies far above the thermal one (e.g. Sironi et al. 2015). 
A much more promising mechanism is the dissipation of magnetic energy through 
reconnection (e.g. Giannios et al. 2009, Uzdensky 2011, Cerutti et al. 2012). 
A prediction of this scenario, resulting from detailed particle--in--cell simulations, 
is the substantial equipartition between the field and the accelerated electrons 
downstream of the reconnection site, where particles cool and emit the radiation we observe (Sironi et al. 2015). Modelling of FSRQ SEDs is in agreement with this prediction (e.g. Ghisellini et al. 2010; Ghisellini \& Tavecchio 2015). Needless to say,  the magnetic reconnection scenario requires that  jets carry a sizeable fraction of their power in the magnetic form up to the emission regions.

As noted above, most of the sources in which the role of magnetic field has been investigated are FSRQs. 
Powerful FSRQs are indeed the majority of the sources belonging to the sample used by Zamaninasab et al. (2014), 
as well as the sources for which equipartition between relativistic electrons and magnetic field is well 
established through the SED modelling. 
By definition, these sources are characterized by the presence of an efficient accretion flow around the central 
BH, flagged by the presence of luminous broad emission lines in the optical spectra or even by the detection 
of the direct emission bump from the accretion disk (e.g. Ghisellini \& Tavecchio 2015). 
In view of the recent insights and problems discussed above it is interesting to extend the investigation to BL Lac objects. Upper limits to the thermal components support the view that the accretion rate in these sources is quite 
low and the accretion flow is likely in the inefficient regime characterizing the advection dominated 
accretion flows (e.g., Ghisellini et al. 2009). In principle, the jet formation and its structure 
could be different to that of FSRQ with powerful and radiatively efficient accretion. 
Moreover, the absence of an important environmental radiation field around the jet implies that the 
high--energy bump in the SED of BL Lac is dominated by the inverse Compton scattering of the synchrotron 
photon themselves (synchrotron self Compton model, SSC). As demonstrated by Tavecchio et al. (1998) 
in this case the relevant physical parameters (most notably for our purposes the magnetic field and 
the electron density) can be uniquely and robustly derived once the SED is relatively well sampled.

In Tavecchio et al. (2010) the SSC model was applied to all the BL Lac detected in the $\gamma$--ray band 
(by Cherenkov telescopes or by {\it Fermi}-LAT in the first three months of operation) and the 
parameters for all sources were derived. 
Here we exploit this large sample to derive the magnetic and the particle energy density of 
BL Lac objects (\S2). 
We anticipate that the magnetic energy density we will derive adopting the popular one--zone 
scenario is quite small compared to that associated to the non-thermal emitting electrons, 
in conflict with the expectations. In \S 3 we review some possible solutions of this problem 
and we discuss our results in \S 4. 

Throughout the paper, the following cosmological  parameters are assumed:
$H_0=70$ km s$^{-1}$ Mpc$^{-1}$, $\Omega_{\rm M}=0.3$, $\Omega_{\Lambda}=0.7$. 
We  use the notation $Q=Q_X \, 10^X $ in cgs units.

\section{Energy densities and powers of BL Lac jets}

The application of simple emission models allows us to derive the basic physical parameters 
of the emission region from the observed SED. 
Specifically, we will apply the one--zone synchrotron--self Compton (SSC) model (e.g. Tavecchio et al. 1998), 
from which we can derive the source size $R$, the Doppler factor $\delta$, the jet comoving frame 
(primed symbols) magnetic field $B^{\prime}$ and the (jet frame) parameters describing the 
broken power law electron energy distribution $N^{\prime}(\gamma)$: the minimum, the break and the 
maximum Lorentz factors $\gamma_{\rm min}$, $\gamma_{\rm b}$ and $\gamma_{\rm max}$, the 
two slopes $n_1$ and $n_2$ and the normalization, $K^{\prime}$.
The assumed phenomenological form of the electron energy distribution is generally valid to reproduce the typical blazar SED bumps, approximately characterized by two power laws with slopes $\alpha_1<1$ and $\alpha_2>1$, connected at the peak frequency.

From these parameters one can directly derive the jet comoving energy density in magnetic field, 
$U^{\prime}_B=B^{\prime \,2}/8\pi$, and in relativistic electrons:
\begin{equation}
U^{\prime}_e=m_{\rm e}c^2\int_{\gamma_{\rm min}}^{\gamma_{\rm max}} 
N^{\prime}(\gamma) (\gamma-1) \, d\gamma \simeq m_{\rm e} c^2 N^{\prime} \langle \gamma \rangle,
\label{ue}
\end{equation}
where $N^{\prime}$ is the total (integrated) electron numerical density and the last expression is 
valid for $\gamma_{\rm min}\gg 1$. 
The corresponding contributions to the jet power (assuming that the emission region 
encompasses the entire jet cross section) $P_{B}$ and $P_{\rm e}$ (e.g. Celotti \& Ghisellini 2008):
\begin{equation}
P_B=\pi R^2 U^{\prime}_B \Gamma^2 \beta c,
\label{pb}
\end{equation}
where the bulk Lorentz factor $\Gamma=\delta$  and:
\begin{equation}
P_{\rm e}=\pi R^2 U^{\prime}_{\rm e} \Gamma^2 \beta c.
\label{pe}
\end{equation}

In the usual case with $n_1=2$ and neglecting the high-energy part of the electron energy distribution, the 
average Lorentz factor is  $\langle \gamma \rangle\simeq \gamma_{\rm min}\ln(\gamma_{\rm b}/\gamma_{\rm min})$. 
Note that, since the average Lorentz factor is typically large, $\langle \gamma \rangle \approx 10^3$, the power 
carried by the electron component is often comparable to that associated by a possible component of cold protons. 
Due to this reason and the uncertainty on the jet composition, we do not include protons in the following and we 
just consider $P_{\rm e}$, noting that this is a {\it strict lower limit} for the power carried by the particle component. 
Note also that the ratio between the magnetic and the electron jet luminosities is just the ratio between the 
corresponding comoving energy densities, $P_{B}/P_{\rm e}=U^{\prime}_{B}/U^{\prime}_{\rm e}$.

\subsection{Analytical estimates: the one zone SSC model}

Along the lines of the analytical treatment in Tavecchio et al. (1998) for the one zone SSC model it 
is possible to derive a useful approximate analytical expression for the ratio $U^{\prime}_{B}/U^{\prime}_{\rm e}$ 
as a function of the observed SED quantities. 
We specialize the following treatment to the case -- usually valid for high-energy emitting BL Lac -- in which 
the SSC peak is produced by scatterings occurring in the Klein-Nishina limit (in any case the detailed numerical 
model include the full treatment of the IC kinematics and cross section).

The magnetic energy density can be evaluated using the synchrotron and the SSC peak frequencies, 
$\nu_{\rm S}$ and $\nu_{\rm C}$. 
In fact, in the KN regime, the observed SSC peak frequency $\nu_{\rm C}$ is related to the energy 
of the electrons at the break $\gamma_{\rm b}m_{\rm e}c^2$ by
\begin{equation}
\nu_{\rm C}=g \gamma_{\rm b} \frac{m_{\rm e}c^2}{h} \delta,
\label{nuckn}
\end{equation}
where $g<1$ is a function of the spectral slopes before and after the peak (see Appendix). Deriving the expression for $\gamma_{\rm b}$ and inserting into the equation for the observed synchrotron peak frequency, $\nu_{\rm S}= B^{\prime} \gamma _{\rm b}^2 \delta e/(2\pi m_{\rm e}c)$, we can express the magnetic energy density using the values of the observed frequencies and the Doppler factor:
\begin{equation}
U^{\prime}_B=\frac{B^{\prime \,2}}{8\pi}=\left( \frac{m_{\rm e}c^2}{h}\right)^4 \frac{g^4 \pi m_{\rm e}^2c^2}{2e^2} \frac{\nu_{\rm S}^2}{\nu_{\rm C}^4} \delta^2
\label{ub}
\end{equation}
Note that $U^{\prime}_B$ depends on the ratio $\nu_{\rm S}^2/\nu_{\rm C}^4$, implying that the typically large 
separation between the two SED peaks results in low magnetic energy densities (see also Tavecchio \& Ghisellini 2008). 
For typical values $\nu_{\rm S}=3\times 10^{16}$ Hz, $\nu_{\rm C}=10^{25}$ Hz and $\delta=15$ we 
derive $ U^{\prime}_B=4\times 10^{-2}$ erg cm$^{-3}$ (we used $g=0.2$, see Appendix).

To calculate $U^{\prime}_{\rm e}$ from Eq. \ref{ue} we need to evaluate the number density of the emitting electrons, 
$N^\prime$. 
To this purpose we start from  the expression for the bolometric observed synchrotron luminosity,
\begin{equation}
L_{\rm S}=\frac{4}{3}\sigma_Tc \, U^{\prime}_B N^\prime \langle \gamma ^2\rangle V^\prime \delta^4
\label{lsyn}
\end{equation}
where $V^\prime=(4/3)\pi R^3$ is the comoving source volume and $\sigma_T$ the Thomson cross section. 
The radius $R$ can be evaluated using the expression for the ratio between the SSC and synchrotron luminosity,
\begin{equation}
\frac{L_{\rm C}}{L_{\rm S}}=\frac{\xi \, U^{\prime}_{\rm S}}{U^{\prime}_{B}},
\label{lratio}
\end{equation}
where $\xi<1$ is a factor accounting for the reduced efficiency of the IC emission in the KN regime 
(see Appendix) and the jet comoving synchrotron radiation energy density is:
\begin{equation}
U^{\prime}_{\rm S} = \frac{L_{\rm S}}{4\pi R^2c\delta^4}.
\label{usyn}
\end{equation}
Combining Eq. \ref{lratio} and Eq. \ref{usyn} we obtain an expression for the radius $R$:
\begin{equation}
R=\left( \frac{\xi L_{\rm S}^2}{4\pi c\delta^4 U_{B} L_{\rm C}} \right)^{1/2}
\label{r}
\end{equation}
Inserting Eq. \ref{r} into the expression for the synchrotron luminosity, Eq. \ref{lsyn}, we obtain:
\begin{equation}
N^\prime=\frac{L_{\rm C}^{3/2}}{L_{\rm S}^2} \frac{9\pi^{1/2}c^{1/2}}
{2\sigma _T \langle \gamma ^2\rangle\, \xi^{3/2}} U^{\prime1/2}_{B}\delta^2
\label{n}
\end{equation}
With these equations at hand we finally derive:
\begin{equation}
{U^{\prime}_B \over U^{\prime}_{\rm e}}= 
\frac{g^2\xi^{3/2}}{1.5\times 10^{-9}}\left( \frac{L_{\rm S}}{L_{\rm C}} \right)^{3/2} L_{\rm S}^{1/2} 
\frac{\langle \gamma ^2\rangle}{\langle \gamma \rangle} \frac{\nu_{\rm S}} {\nu_{\rm C}^2}\, \delta^{-1}
\label{ratiofin}
\end{equation}
%

\begin{figure}
\vspace*{-1. truecm}
\hspace*{-0.8 truecm}
\psfig{file=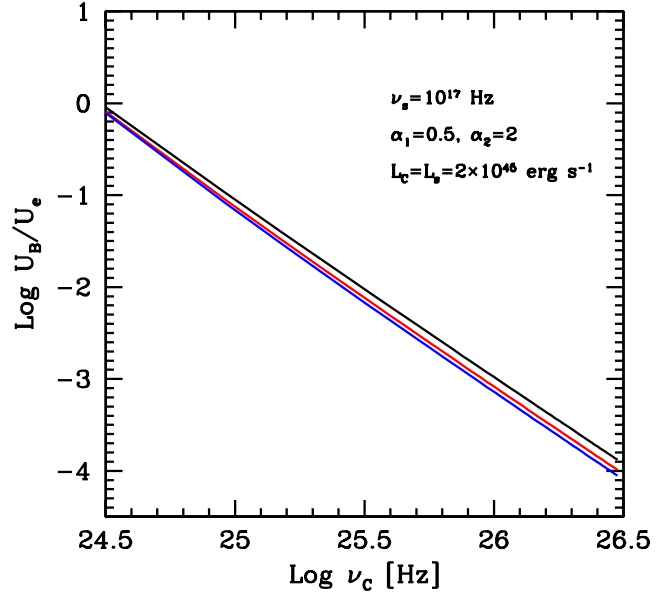,height=10.5cm,width=10.5cm}
\vspace{-1.1 cm}
\caption{
Ratio between the jet frame magnetic and the relativistic electron energy density as a function of 
the observed SSC peak frequency, assuming typical values for the other quantities (indicated in the legend). 
The three curves corresponds to different values of the Doppler factor, $\delta=10$ (black), 20 (red) and 30 (blue).
}
\label{equi}
\end{figure}

For the assumed broken power law slopes of the electron energy distribution, 
only the first branch $\gamma<\gamma_{\rm b}$ is relevant for the calculation of the 
ratio $\langle \gamma ^2\rangle / \langle \gamma \rangle$. 
The typical low energy synchrotron spectral slope is $\alpha_1=0.5$, corresponding to $n_1=2$.
Using this value we obtain  
$\langle \gamma ^2\rangle / \langle \gamma \rangle=\gamma_{\rm b}/\ln (\gamma_{\rm b}/\gamma_{\rm min})$
($\gamma_{\rm b}$ can be obtained through observed quantities using Eq. \ref{nuckn}).

Typical values for TeV BL Lacs are $\nu_{\rm S}=10^{17}$ Hz, $\nu_{\rm C}=3\times 10^{25}$ Hz, 
$L_{\rm S}=2\times 10^{45}$ erg s$^{-1}$, $L_{\rm S}/L_{\rm C}\simeq 1$, $\delta=15$, 
$n_1=2$.
Using these values, the ratio $U^{\prime}_{B}/U^{\prime}_{\rm e}$ turns out to be 
much smaller than unity: $U^{\prime}_{B}/U^{\prime}_{\rm e}\simeq 10^{-2}$. 

Fig. \ref{equi} shows the ratio $U^{\prime}_{B}/U^{\prime}_{\rm e}$ from Eq. \ref{ratiofin} 
as a function of the SSC peak frequency $\nu_{\rm C}$ for typical values of the other parameters and 
three values of the Doppler factor, $\delta=10, 20$ and 30. 
For the assumed $\alpha_1=0.5$, $U^{\prime}_{B}/U^{\prime}_{\rm e}$ is only weakly dependent on the 
Doppler factor ($U^{\prime}_{B}/U^{\prime}_{\rm e} \propto \delta^{-1/2}$), while it depends quite 
strongly on the SSC peak frequency, $U^{\prime}_{B}/U^{\prime}_{\rm e}\propto \nu_{\rm C}^{-7/4}$. 

This approximate calculation shows that with a good sampled SED (especially around the peaks) in the 
one--zone framework the ratio $U^{\prime}_{B}/U^{\prime}_{\rm e}$ is robustly and uniquely derived from the observed quantities, with a weak dependence on the value of the Doppler factor. This is clearly a consequence of the fact that, as stressed in Tavecchio et al. (1998), the parameters specifying the one-zone SSC model are uniquely determined once the basic observables (i.e. the synchrotron and the SSC peak frequencies and luminosities plus the variability timescale) are known. In the previous derivation we do not make use of the causality relations connecting the radius and the Doppler factor to the observed minimum variability timescale and thus the Doppler factor still explicitly appears in Eq. \ref{ratiofin}.

\begin{figure}
\vspace*{-1. truecm}
\hspace*{-0.8 truecm}
\psfig{file=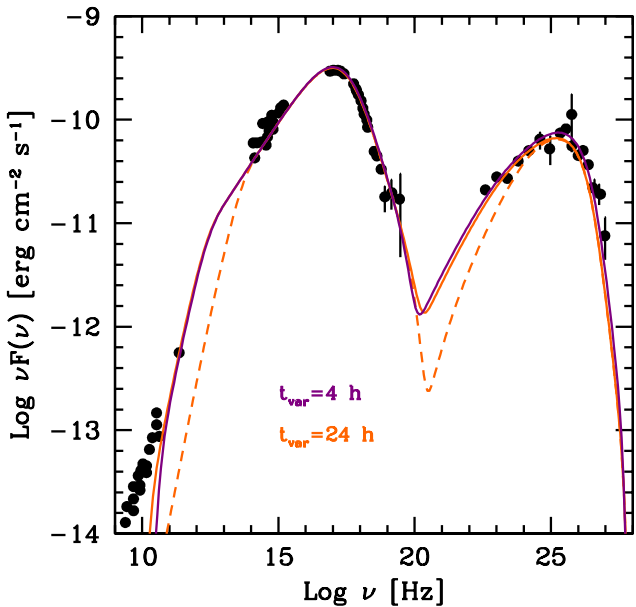,height=10.5cm,width=10.5cm}
\vspace{-1.3 cm}
\caption{Spectral energy distribution of Mkn 421 (black filled circles) obtained during the campaign reported in Abdo et al. (2011). The violet and orange solid lines show the theoretical SED calculated with the SSC model for two different value of the variability timescale with the parameters reported in Table 2 (model 1 and 2 respectively). The dashed orange line is for model 2 assuming the larger $\gamma_{\rm min}$ allowed by the data, $\gamma_{\rm min}=3\times 10^3$.
}
\label{421}
\end{figure}

\begin{table*}
\centering
\begin{tabular}{cccccccccccccc}
\hline
\hline
Model    & $\gamma _{\rm min}$ & $\gamma _{\rm b}$& $\gamma _{\rm max}$& $n_1$&$n_2$ &$B$ &$K$ &$R$ 
& $\delta $ & $U_B$ & $U_{\rm e}$ & $P_B$ & $P_{\rm e}$\\
\quad [1] & [2]  & [3] & [4] & [5] & [6] & [7] & [8]  & [9] & [10]  & [11] & [12] & [13] & [14]\\
\hline
1    &$500 $&$ 1.7\times 10^{5} $&$ 2\times 10^{6} $&$ 2.2 $&$ 4.8 $&$ 0.075 $&$ 1.3\times 10^{4}$&$ 1 $& 25  & $2.2\times 10^{-4}$& $1.1\times 10^{-2}$ & 0.13 & 9.0\\      
2    &$700 $&$ 2.5\times 10^{5} $&$ 4\times 10^{6} $&$ 2.2 $&$ 4.8 $&$ 0.06 $&$ 3.2\times 10^{3}$&$ 3.6 $& 14 & $1.4\times 10^{-4}$& $2.6\times 10^{-3}$ & 0.34 & 8.5\\   
\hline
S&$400 $&$ 1.2\times 10^{5} $&$ 10^{6} $&$ 2.2 $&$ 4.9 $&$ 0.18 $&$ 1.3\times 10^{3}$&$ 1 $& 30  & $1.2\times 10^{-3}$& $1.2\times 10^{-3}$ & 1.1 & 1.1\\    
L&$10 $&$ 1.5\times 10^{5} $&$ 10^{7} $&$ 2. $&$ 4.0 $&$ 0.32 $&$ 2.8\times 10^{1}$&$ 2 $& 5  & $4\times 10^{-3}$& $2.3\times 10^{-4}$ & 0.1 & $8\times 10^{-3}$\\    
\hline
\hline
\end{tabular}
\vskip 0.4 true cm
\caption{
Input model parameters for the models of Mkn 421 in Figs. \ref{421} (first two raws) 
and \ref{421spinelayer} (second trwo raws) and derived magnetic and electron power. 
[1]: model.  
[2], [3] and [4]: minimum, break and maximum electron Lorentz factor.  
[5] and [6]: slope of the electron energy distribution below and above $\gamma _b$. 
[7]: magnetic field [G]. 
[8]: normalization of the electron distribution in units of cm$^{-3}$. 
[9]: radius of the emission zone in units of $10^{16}$ cm. 
[10]: Doppler factor. 
[11]: Magnetic energy density, erg cm$^{-3}$. 
[12]: Relativistic electron energy density, erg cm$^{-3}$.
[13]: Poynting flux carried by the jet, in units of $10^{43}$ erg s$^{-1}$. 
[14]: kinetic power carried by the emitting electrons of the jet, in units of $10^{43}$ erg s$^{-1}$. 
For the spine--layer model, the third and the fourth raws report the parameters for the spine (S) 
and the layer (L), respectively. For the layer $\Gamma=2.6$ is assumed.}
\label{tableparam}
\end{table*}

\subsection{An illustrative case: Mkn 421}

Before presenting the results for the entire sample, it is worth to discuss a specific case, 
showing how the SED fitting method allows one to obtain quite robust estimates of the physical quantities. 
To this purpose we focus our attention on Mkn 421, one of the BL Lac detected at TeV energies 
characterized by the most complete SED coverage. 
In particular, Abdo et al. (2011) presented a SED with a nicely complete sampling from radio up to 
TeV $\gamma$--rays obtained during an intensive multifrequency campaign. 
The source was in a relatively quiescent state, thus representing a sort of average activity state of the jet. 
The SED, with two possible realizations of the SSC model, is reported in Fig. \ref{421}.

As already stressed, the one-zone SSC model parameters are uniquely specified once the SED bumps (namely, 
peak frequencies and luminosities) and the variability timescale are well characterized (Tavecchio et al. 1998). 
The latter observable determines the value of the source size, through the causality relation 
$R\approx c t_{\rm var}\delta/(1+z)$. 
The values of the peak frequencies and luminosities can be linked to the other physical parameters, 
most notably the magnetic field and the particle density and energy.

The available data (Fig. \ref{421}) provide an excellent description of both the synchrotron and the IC peak.
Less constrained is the variability timescale. 
Indeed the time coverage of the source during the campaign was not suited to probe short 
variability and it only allows Abdo et al. (2011) to follow daily--scale variations. 
Activity at shorter (even to sub--hour, e.g. Alekisc et al. 2011) timescale cannot however be excluded. 
For this reason we performed two fits of the SED, with $t_{\rm var}=4$ h and 24 h. 
The parameters used to reproduce the SED (with the model described in Maraschi \& Tavecchio 2003), 
together with the energy densities and the jet powers derived using Eq. \ref{pb}--\ref{pe}, 
are reported in Table 1. 

In the model, the electron energy distribution is phenomenologically assumed to follow a broken power law shape, with the break at the energy $\gamma_{\rm b } m_ec^2$. We remark that the distribution is fixed only by the condition to reproduce the observed SED, without taking self-consistently into account the evolution due to injection and cooling effects (e.g. Kirk et al. 1998).

As expected after the discussion above, it is clear that, although some of the parameters are different 
in the two cases (in particular the source radius), the ratio between the energy densities is 
relatively stable, implying that the energy density (and the power) associated to the
non--thermal electrons  is largely (by a factor 20--50) dominant over that of the magnetic field 
(see the similar conclusion in Abdo et al. 2011 for the same dataset). 
The largest ratio is obtained for the case of the shorter (and more usual) variability timescale. 
This result is rather general: the shorter the variability timescale the smaller the derived 
magnetic field (see also Aleksic et al. 2011). 
 
The reason for the small magnetic field (and relatively large Doppler factor) required by 
the SSC modeling of TeV emitting sources can be traced back to the large separation between the two SED peaks. 
Indeed, to produce the high--energy peak at $\nu_{\rm C}\sim 10^{25}$ Hz the break energy of the 
electron distribution should be at least of $\gamma_{\rm b}> h \nu_{\rm C}/\delta \, m_{\rm e}c^2\approx 10^5$. 
Since the synchrotron bump peaks around the soft X-ray band, $\nu_{\rm S}\approx 10^{17}$  Hz, 
we directly derive a typical magnetic field of $B \lesssim 0.3 \, \delta_1^{-1}$. 

A limited possibility to reduce the $U^{\prime}_{\rm e}/U^{\prime}_B$ ratio is by decreasing the number 
of radiating electrons, increasing the minimum Lorentz factor, $\gamma_{\rm min}$.  
An upper limit to $\gamma_{\rm min}$ is however provided by the condition to reproduce the optical and the 
{\it Fermi}--LAT data, tracking the low--energy part of the synchrotron and IC bump, respectively. 
An example is given in Fig. 1, for which $\gamma_{\rm min}$ for Model 2 has been increased to 
$\gamma_{\rm min}=3\times 10^3$ (dashed orange line). 
Even in this extreme case the electron energy density decreased only slightly, from 
$U_{\rm e}=2.6\times 10^{-3}$ erg cm$^{-3}$ to $U_{\rm e}=1.7\times 10^{-3}$ erg cm$^{-3}$.

In passing, we also note that the cooling time of the electrons at the break (i.e. those emitting at the peak) is comparable with the dynamical timescale $t_{\rm dyn }\approx R/c$, consistently with the idea that the break in the electron energy distribution is related to  radiative losses.

This example confirms that the estimate of the magnetic and kinetic (electronic) energy density 
(and the associated powers) in the framework of the one--zone model is quite robust and reliable. 

\begin{figure}
\vspace*{-1.2 truecm}
\hspace*{-0.8 truecm}
\psfig{file=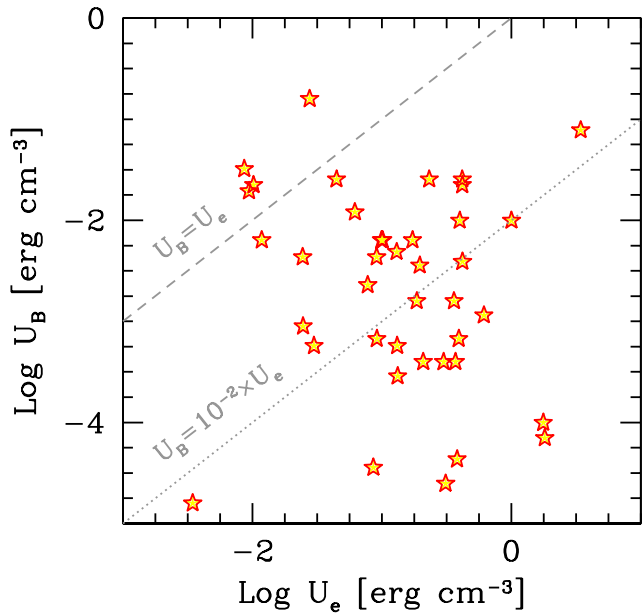,height=10.5cm,width=10.5cm}
\vspace{-1.1 cm}
\caption{
Magnetic energy density ($y$--axis) and relativistic electron energy density ($x$--axis) 
derived using the physical parameters inferred through the modeling of the SED by 
Tavecchio et al. (2010). The grey dashed line shows the equality $U_{\rm e}=U_{B}$. 
The great majority of the sources occupy the region $U_{\rm e}\gg U_{B}$.
}
\label{magndens}
\end{figure}

\begin{figure}
\vspace*{-1.2 truecm}
\hspace*{-0.8 truecm}
\psfig{file=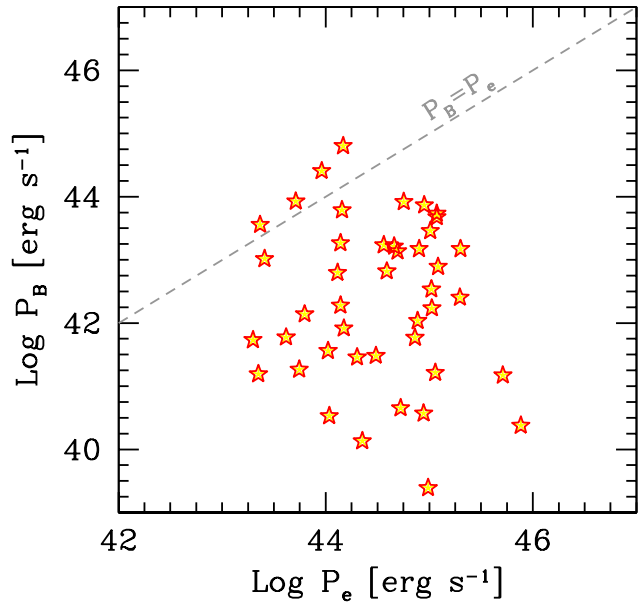,height=10.5cm,width=10.5cm}
\vspace{-1.1 cm}
\caption{
Power carried by BL Lac jets in the form of magnetic field ($y$--axis) and relativistic electrons 
($x$--axis) derived using the physical parameters inferred through the modeling of the SED by 
Tavecchio et al. (2010). The gray dashed line shows the equality $P_{\rm e}=P_{B}$. 
The great majority of the sources occupy the region $P_{\rm e}\gg P_{B}$.}
\label{magn}
\end{figure}

\subsection{The full sample}

Tavecchio et al. (2010) obtained the physical parameters associated to 45 BL Lacs applying the 
one--zone synchrotron-self Compton (SSC) model (e.g. Tavecchio et al. 1998) to non-simultaneous SED, 
whose IC peak is tracked by either {\it Fermi}--LAT (33 sources) or TeV (12 sources) data. 

The magnetic and the electron energy densities derived using the parameter values provided in 
Tavecchio et al. (2010) are reported and compared in Fig. \ref{magndens}. 
As already discussed for the case of Mkn 421, most of the sources are characterized by an electron 
component strongly dominating over the magnetic one, with an average ratio 
$U^{\prime}_{\rm e}/U^{\prime}_{B}\sim 100$. 
As discussed above, the uncertainty affecting both $U^{\prime}_{\rm e}$ and $/U^{\prime}_{B}$ 
is limited, and cannot account for a systematic error of such a large ratio. 
In Fig. \ref{magn} we also report the comparison between the corresponding powers. 
Again, the magnetic power is largely below that associated to the relativistic electrons 
in most of the sources. 
The non--simultaneity of several SED and the quality of the data that is not always optimal impacts 
on the derived parameters for single objects, but this is compensated by the large number of sources 
ensuring that, from the statistical point of view, the estimates can be trusted.

\begin{figure}
\vspace*{-1.2 truecm}
\hspace*{-0.8 truecm}
\psfig{file=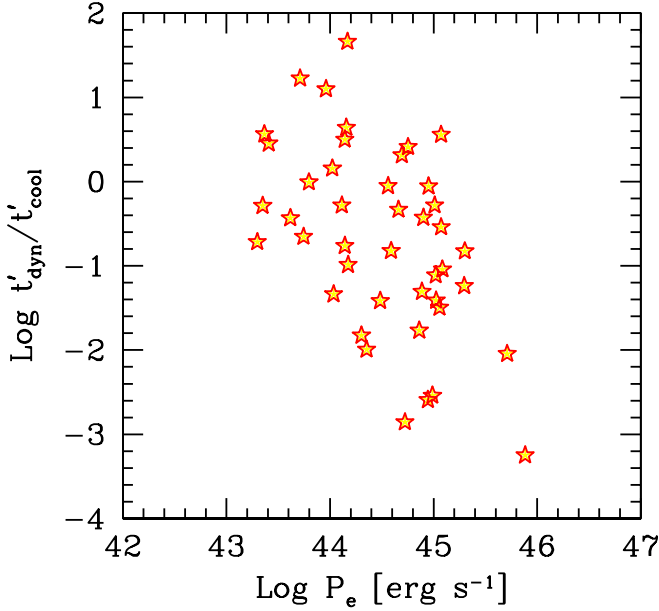,height=10.5cm,width=10.5cm}
\vspace{-1.1 cm}
\caption{Ratio of the dynamical and cooling times (for electrons emitting at the SED peak) as a function of the electron power carried by the jet. Most of the sources lies in the region $t^{\prime}_{\rm dyn}/t^{\prime}_{\rm cool}<1$, implying a small radiative efficiency.
}
\label{effic}
\end{figure}

We also note that the small value of the magnetic energy density characterizing the BL Lacs of our sample --  related to the level of the synchrotron (and SSC) emission -- has  an impact on the radiative efficiency of the jets. In Fig. \ref{effic} we  report the value of the ratio between the dynamical timescale $t^{\prime}_{\rm dyn}\simeq R/c$ and $t^{\prime}_{\rm cool}$, the cooling time of the electrons with Lorentz factor $\gamma_{\rm b}$ -- i.e. those emitting at the SED peak -- as a function of the jet power carried by the relativistic electrons, $P_{\rm e}$. Clearly, the majority of the sources lies in the region with $t^{\prime}_{\rm dyn}<t^{\prime}_{\rm cool}$, implying a small radiative efficiency for most these jets (note that for Mkn 421 we obtained $t^{\prime}_{\rm dyn}\approx t^{\prime}_{\rm cool}$). The low efficiency is clearly related to the small magnetic field. We can conclude that the simple one-zone leptonic model for BL Lacs foresees quite inefficient jets, implying large jet powers.  Note also that, since the cooling time for the electrons emitting at the synchrotron peak is smaller than the typical jet dynamical timescale, the break in the electron energy distribution required to reproduce the SED shape cannot be related to the radiative losses of the electrons (as for the case of FSRQ, e.g. Ghisellini et al. 2010). In fact a possible cooling break should appear at much larger frequencies -- for which the relations $t^{\prime}_{\rm dyn}\simeq t^{\prime}_{\rm cool}(\gamma)$ holds for the electrons with Lorentz factor $\gamma$--, i.e. in the hard X-ray band, which is not well sampled by currents observations. 

\section{Possible alternatives}

The results of the preceding section show that in the framework of the one--zone 
SSC model for BL Lac jets, the energy density associated to the  relativistic electron component 
largely exceeds that of the magnetic field. 
Given the striking difference with the case of FSRQ and the conflict with expectations from the 
theoretical scenario, it is compelling to investigate whether this conclusion can be revised using 
a different setup of the emission model. 
Needless to say, the most direct possibility is to relax the strong assumption that the emission 
involves only one homogeneous region. 
This line is supported by growing evidence that the emission sites could indeed be more  complex 
than what is usually depicted in the simple one--zone scheme. 
In the following we discuss two possible alternative scenarios that -- at a first sight -- could provide 
a solution to the problem.

\subsection{Magnetic reconnection in compact sites}

As discussed in the Introduction, there is growing evidence that processes triggered by magnetic reconnection events could provide an effective way to account for particle acceleration in relativistic outflows. Support to the idea that emission in blazars can be associated to reconnection sites comes from independent arguments based i) on the observed ultra--fast variability (e.g. Giannios et al. 2009, 
Giannios 2013, Nalewajko et al. 2011) and ii) on the inefficient acceleration provided by shocks (Sironi et al. 2015).  
In this context one thus expects that the emission from blazars (or at least a part of it) is produced by magnetic reconnection in compact regions. 
However, as discussed above, assuming that the entire emission is produced by compact regions does 
not solve the problem posed by our results, since, as long as the synchrotron and the IC (SSC) 
emission are produced by the same region(s), the magnetic--to--electron energy density ratio is uniquely fixed by the SED at the values derived in \S 2.  

A possibility is to decouple the emission, assuming that the compact regions are responsible 
for the (energetic and rapidly variable) high--energy component, while the low--energy radiation is 
produced by electrons living in a larger volume of the jet. 
This scheme would mimic scenarios in which the emission zone comprises one (or more) reconnection island, 
in rough equipartition, and a larger, magnetically dominated region -- the jet --, 
where particles escaping from the islands diffuse and radiatively cool (e.g. Nalewajko et al. 2011). 
The emission from such a system would be characterized by a strong synchrotron component, 
produced in the large magnetized region by cooled electrons, and by a IC component produced in 
the magnetic islands by freshly accelerated electrons. 

\begin{figure}
\vspace*{-1. truecm}
\hspace*{-1.1 truecm}
\psfig{file=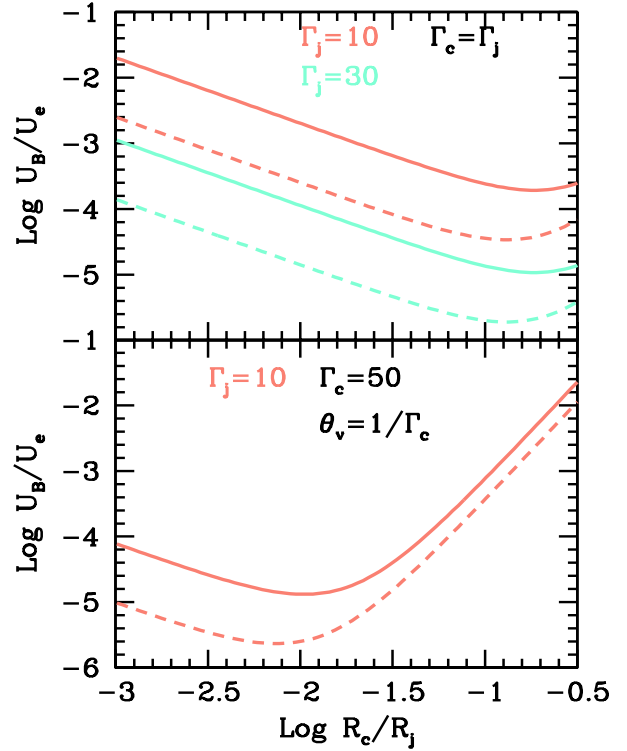,height=11cm,width=11.5cm}
\caption{Ratio of magnetic and electronic energy density in the model assuming one compact region embedded in the jet, as a function of the ratio between the compact region radius and that of the jet. In the upper panel we assume that the compact region is characterized by the same bulk Lorentz factor of the jet. The curves reports the result for $\delta=10$ and 30 and for $L_{\rm S,c}=2\times10^{44}$ erg s$^{-1}$ (solid) and $10^{44}$ erg s$^{-1}$ (dashed). In the lower panel we assume a larger bulk Lorentz factor for the compact emitting region, $\Gamma_{\rm b}=30$.
}
\label{reconn}
\end{figure}

In this set-up, we can use the analogue of Eq. \ref{lratio} to fix the magnetic energy density in the compact region (double primes reference frame):
\begin{equation}
\frac{L_{\rm C,c}}{L_{\rm S,c}}=\frac{\xi U^{\prime \prime}_{\rm rad}}{U^{\prime \prime}_B} \; \rightarrow \; U^{\prime \prime}_B = \xi U^{\prime \prime}_{\rm rad} \frac{L_{\rm S,c}}{L_{\rm C,c}}
\end{equation}
where $L_{\rm S,c}$ and $L_{\rm C,c}=L_{\rm C}$ are the observer frame luminosity of the synchrotron and IC emission of the compact region, respectively, and $U^{\prime \prime}_{\rm rad} $ is the sum of the density of the radiation produced in the jet and that locally produced in the compact region, as measured in the compact region reference frame:
\begin{equation}
U^{\prime \prime}_{\rm rad} = U^{\prime }_{\rm j} \Gamma_{\rm rel}^2 + U^{\prime \prime}_{\rm c}
\end{equation}
(where the subscript "c" stays for "compact"). The factor $\Gamma_{\rm rel}^2$ takes into account the boosting of the jet radiation energy density in the compact region rest frame, if the latter is characterized by relativistic speeds in the jet frame. The relative Lorentz factor is given by $\Gamma_{\rm rel}=\Gamma_{\rm j}\Gamma_{\rm c}(1-\beta_{\rm j}\beta_{\rm c})$.

Using the analogue of Eq. \ref{usyn} and rearranging the terms we obtain:
\begin{equation}
U^{\prime \prime}_{\rm rad} = \frac{L_{\rm S}}{4\pi R_{\rm j}^2 c \delta_{\rm j}^4} \left[ \Gamma^2_{\rm rel} + \frac{L_{\rm S,c}}{L_{\rm S}} \left(\frac{R_{\rm j}}{R_{\rm c}}\right)^2 \left(\frac{\delta_{\rm j}}{\delta_{\rm c}}\right)^4 \right]
\label{uradblob}
\end{equation}
The electron density (to be used in Eq. \ref{ue} to derive $U^{\prime \prime}_{\rm e}$) can be derived from the expression for the IC luminosity:
\begin{equation}
L_{\rm C,c}=\frac{4}{3}\sigma_Tc \, U^{\prime\prime}_{\rm rad} N^{\prime \prime}\langle \gamma ^2\rangle V^{\prime \prime} \delta _{\rm c}^4.
\label{lcblob}
\end{equation}

By construction $L_{\rm S,c}\ll L_{\rm S,j}$ (since the observed synchrotron component is dominated by the emission from the large region). Note that in the present case, in which the observed emission comprises the contribution from two (or more) independent regions, not all the parameters are fixed by the SED as in the one-zone case discussed before.

In Fig. \ref{reconn} we show the derived value of the ratio of the magnetic and electron energy densities as a function of the ratio of radii of the compact emission region and the jet, $R_{\rm c}/R_{\rm j}$, assuming $L_{\rm S}=L_{\rm C}=2\times10^{45}$ erg s$^{-1}$, $R_{\rm j}=3\times 10^{15}$ cm.
In the upper panel we show the case in which the compact region is not relativistically moving in the jet frame, for two values of the bulk Lorentz factors ($\Gamma_{\rm j}=10$ and 30) and two values of the luminosity of the synchrotron emission of the compact region, $L_{\rm S,c}=2\times10^{44}$ erg s$^{-1}$ and $10^{44}$ erg s$^{-1}$ (i.e. 10\% and 5\% of the observed synchrotron peak luminosity). In the lower panel we assume instead that the compact region has a Lorentz factor $\Gamma_{\rm c}=50$ (as measured in the observer frame) and the jet $\Gamma_{\rm j}=10$. In all cases we assume that the system is observed under the most favorable angle for the compact region, $\theta_{\rm v}=1/\Gamma_{\rm c}$.

The existence of a minimum of  $U^{\prime \prime}_B/U^{\prime \prime}_{\rm e} $ as a function of $R_{\rm c}$ -- visible in Fig. \ref{reconn} -- can be explained as follows. 
Let us start with a relatively large compact region, so that the radiation energy density of the jet (as measured in the frame of  the compact region) is larger than that produced by the compact region itself, $U^{\prime }_{\rm j} \Gamma_{\rm rel}^2 > U^{\prime \prime}_{\rm c}$ (see Eq. \ref{uradblob}). In this case, since the radiation energy density is fixed, to produce a fixed IC luminosity, the system requires a fixed number of electrons. Furthermore, since the synchrotron luminosity that the system has to produce is also fixed, also the magnetic energy density is fixed. Considering now a smaller radius of the source (but large enough so that the jet radiation energy density still dominates), the only quantity that changes is the electron density, since the same amount of electrons is confined in a smaller volume. This directly implies that the ratio 
 $U^{\prime \prime}_B/U^{\prime \prime}_{\rm e} $ decreases with decreasing $R_{\rm c}$. For $R_{\rm c}$ small enough, instead, the local energy density dominates. In this regime, decreasing $R_{\rm c}$ implies the increase of the soft photon energy density (since the constant synchrotron luminosity is produced in a more compact region), leading to the reduction of the number of electrons (and thus of the electron energy density). To keep the synchrotron luminosity constant, the magnetic energy density has to increase. The combination of these two effects leads to increase  $U^{\prime \prime}_B/U^{\prime \prime}_{\rm e} $. The minimum of $U^{\prime \prime}_B/U^{\prime \prime}_{\rm e}$ is therefore located to the radius where $U^{\prime }_{\rm j} \Gamma_{\rm rel}^2 = U^{\prime \prime}_{\rm c}$

Fig.\ref{uradblob} clearly shows that in all cases the system is still strongly unbalanced toward the electron component. We conclude that even in this framework one cannot reproduce the equipartition conditions.

\subsection{Structured jet/external soft photon component}

A second possibility that could alleviate the problem is to partly decouple the synchrotron and IC components, 
assuming the existence of a supplementary source of soft photons intervening in the IC emission. 
Note indeed that, according to Eq. \ref{lratio}, for a constant ratio of IC and synchrotron luminosities, 
a larger radiation energy density -- implying a reduced number of emitting electrons -- 
allows a larger magnetic field energy density (for a  constant synchrotron output). 
Such a scheme is naturally implemented in the so--called structured (or spine--layer) jet scenario 
(Ghisellini, Tavecchio \& Chiaberge 2005), inspired by peculiar features of TeV emitting BL Lacs, 
such as the absence of fast superluminal components, the presence of a edge--brightened radio 
structure and the issues related to the unification with radiogalaxies. 
In this model one assumes the existence of two regions in the jet: a faster inner core (the spine), 
surrounded by a slower sheath of material (the layer). 
Given the amplification of the radiation emitted by one region as observed in the frame of the 
other caused by the relative motion, the IC luminosity of both components (in particular that of 
the spine) is increased with respect to that of the one--zone model. 
The emission from blazars is dominated by the spine, and the less beamed layer 
emission is thought to be visible only for misaligned jets (e.g. Sbarrato et al. 2014).

To investigate the possibility to increase the magnetic--to--electron energy density in this 
framework, it is useful to derive some simple analytical relations that can help us to fix ideas. 
Working again in the KN regime, note that Eq. \ref{nuckn}-\ref{ub} -- which do not involve any 
information about the soft photon field -- applies also in the present case.
Eq. \ref{lsyn} expressing the synchrotron luminosity can be used to derive the electron number 
density and thus the electron energy density:
\begin{equation}
U^{\prime}_{\rm e}=\frac{9L_sm_e c}{16\pi \sigma_T R^3U^{\prime}_{B} \delta^4} \frac{\langle \gamma \rangle}{\langle \gamma^2 \rangle}.
\end{equation}
Using Eq. \ref{ub} for $U^{\prime}_B$ and considering only the relevant parameters we get:
\begin{equation}
\frac{U^{\prime}_B}{U^{\prime}_{\rm e}}  \propto \frac{R^3 \delta^8}{L_s} \frac{\nu_s^4} {\nu_C^8}\frac{\langle \gamma^2 \rangle}{\langle \gamma \rangle}.
\end{equation}

Note that, since now the level of the IC luminosity can be tuned acting on the level of the layer external 
radiation field, we cannot -- as in the case of one-zone model -- completely fix all the parameters.
In particular the strong dependence on $R$ and $\delta$ suggests that with a relatively large value of 
the Doppler factor the model can achieve equipartition. 
Note that in the previous arguments we did not discuss the characteristics of the layer emission, 
which is thought to be able to provide the required level of soft radiation. 
The only strong constraints that can be put is that the level of the synchrotron component of the layer, 
as observed at Earth, is much less than the spine synchrotron emission.

\begin{figure}
\vspace*{-1. truecm}
\hspace*{-0.8 truecm}
\psfig{file=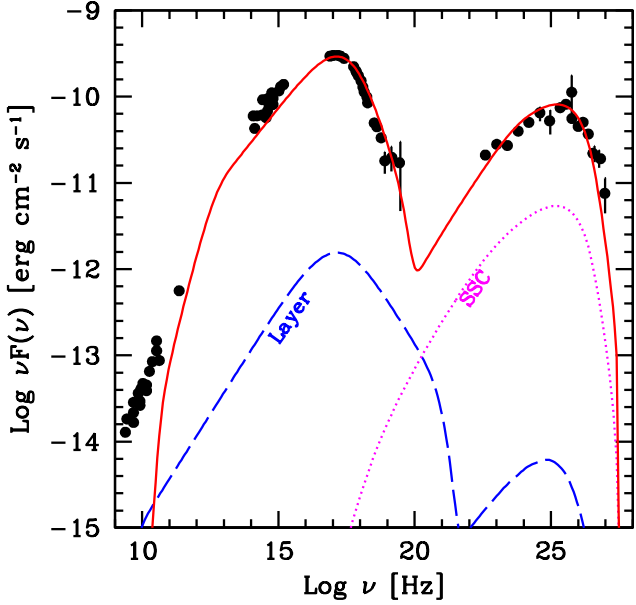,height=10.5cm,width=10.5cm}
\vspace{-1.3 cm}
\caption{As Fig. \ref{421}. 
Lines show the results of  the structured jet model. 
The dashed blue line shows the emission from the layer. 
The magenta dotted line reports the SSC emission from the spine. 
See text for details.
}
\label{421spinelayer}
\end{figure}

In Fig. \ref{421spinelayer} we report, for the specific case of Mkn 421, a possible modelization of the 
SED with the spine-layer model (parameters are reported in Table 1). 
The parameters have been found trying to reproduce at best the SED keeping the system at equipartition. 
As suggested above, this can be achieved with a relatively large Doppler factor, $\delta=30$. 
We remark again that, as long as the scattering between the spine electrons and the layer seed photons occurs in the KN regime, only the luminosity of the layer soft emission is important, while the details of its spectrum do not affect the spine high energy emission.
We conclude that the structured jet model can satisfactorily reproduce the SED in equipartition conditions 
with a reasonable choice of the parameters.

\section{Conclusions}

The current general scenario for jet production assumes that jet starts as a strongly magnetized flow  
and remains highly magnetized up to large distances. 
A highly magnetized plasma is also required by magnetic reconnection models, according to 
which the non--thermal particle population is energized at the magnetic reconnection sites. 
Recent particle-in-cells simulations (Sironi et al. 2014) further show that the post--reconnection downstream regions should be characterized 
by a substantial equipartition between particles and magnetic fields.

BL Lac objects represent the best systems to investigate the role of magnetic field. 
In standard one--zone models all parameters are fixed by current observations of the BL Lac SED and 
the jet comoving energy densities are robustly determined.
We have shown that in this framework the magnetic energy densities comes out to be 1--2 order of magnitude smaller that that associated to relativistic electrons. Occasionally, some works in the past reported the same conclusion on single sources (e.g. Abdo et al. 2009, Acciari et al. 2011, Aleksic et al. 2012, 2015), but we have demonstrated that this is a common property of the large majority of BL Lacs, stemming from the typical SED parameters.

We have shown that a viable possibility to reconcile the observations with the theoretical scenario 
is to relax the assumption that the emission involves only one region. 
In the structured jet  model (Ghisellini et al. 2015), in which the seed photons for the emission 
from the spine are mainly provided by the surrounding slow layer, it is indeed possible to reproduce 
the observed SED assuming equipartition between the magnetic and the electron energy densities.
We recall here that the existence of the postulated structure in BL Lac jets (and FRI radiogalaxies) 
is supported by a variety of observational and phenomenological arguments 
(e.g. Chiaberge et al. 2000, Giroletti et al. 2004, Piner \& Edwards 2014, Nagai et al. 2014)
and also by numerical simulations (e.g. Rossi et al. 2008). We have also suggested that structured BL Lac jets could provide the ideal environment to produce the PeV neutrinos detected by IceCube (Tavecchio et al. 2014).

Another -- though less attractive -- possibility would be to assume that the emission involves only a 
single region of the jet (and thus would be satisfactorily described by one--zone models) 
but the inferred  magnetic fields are not representative of the actual fields carried by the jet. 
To satisfy the (conservative) conditions $U^{\prime}_{B}\approx U^{\prime}_{\rm e}$ (and $P_{B}\approx P_{\rm e}$) 
one requires magnetic energy densities larger by a factor 10--100 than those derived by the 
SED modeling with one--zone models. 
A somewhat {\it ad hoc} possibility (see also Nalewajko et al. 2014) is to assume that while 
the observed emission is produced in a matter dominated core of the outflow (the {\it spine}), 
there is a magnetically dominated layer whose associated magnetic luminosity satisfy the above requirements. Along the same lines one could argue that the emission is limited to localized compact sites 
embedded in a much magnetized plasma.

\section*{Acknowledgments}
We thank a referee for useful comments. This work has been partly founded by a PRIN-INAF 2014 grant. We thank Lorenzo Sironi for stimulating discussions.

\appendix
\section{}

The parameter $g$ is derived in Tavecchio et al. (1998) assuming a step function approximation for the Klein-Nishina cross-section. Assuming that $\alpha _1$ and $\alpha _2$ are the spectral slopes of the synchrotron bump before and above the peak, its expression is:
\begin{equation}
g(\alpha_1,\alpha_2)=\exp \left[ \frac{1}{\alpha_1-1} + \frac{1}{2(\alpha_2-\alpha_1)} \right].
\end{equation}
For typical values $\alpha_1$=0.5, $\alpha_2=2$, $g=0.18$.
\\

\noindent
The parameter $\xi$ in Eq. \ref{lratio}:

\begin{equation}
\frac{L_{\rm C}}{L_{\rm S}}=\frac{\xi \, U^{\prime}_{\rm s}}{U^{\prime}_{B}},
\end{equation}
takes into account the fact that in KN regime the SSC output is suppressed with respect to the Thomson case (for which $\xi=1$). In Tavecchio et al. (1998) an approximate value is derived which accounts for the reduction of the frequencies of the available target photons  but does not include the fact that the electrons emitting at the SSC peak have a Lorentz factor $\gamma_{\rm KN}= g \gamma_{\rm b} <\gamma_{\rm b}$. Here we derive the complete expression.

The ratio $L_{\rm C}/L_{\rm S}$ can be expressed as:
\begin{equation}
\frac{L_{\rm C}}{L_{\rm S}}= \frac{U^{\prime}_{\rm s,avail}N(\gamma_{\rm KN})\gamma^3_{\rm KN}}{U^{\prime}_{B} N(\gamma_{\rm b})\gamma^3_{\rm b}}
\label{ratiokn}
\end{equation}
where $U^{\prime}_{\rm s,avail}$ is the energy density available for scattering with electrons of energy $\gamma_{\rm KN}$, which using the step function KN approximation gives (Tavecchio et al. 1998):
\begin{equation}
U^{\prime}_{\rm s,avail}=U^{\prime}_{\rm s} \left( \frac{3m_{\rm e}c^2\delta}{4h\gamma_{\rm KN}\nu_{\rm S}}\right)^{1-\alpha_1},
\end{equation}
where $\gamma_{\rm KN}$ can be expressed through the SSC peak frequency (see \S2.1), $\gamma_{\rm KN}=h\nu_{\rm C}/m_{\rm e}c^2\delta$. Recalling that the electrons follow a power law energy distribution with slope $n_1=2\alpha_1+1$ (the high energy tail above $\gamma_{\rm b}$ is unimportant here), Eq. \ref{ratiokn} can be finally written as:
\begin{equation}
\frac{L_{\rm C}}{L_{\rm S}}= \frac{U^{\prime}_{\rm s}}{U^{\prime}_{B}} \left[ \frac{3}{4} \left( \frac{m_{\rm e}c^2}{h}\right)^2 \frac{\delta^2}{\nu_{\rm S}\nu_{\rm C}}  \right]^{1-\alpha_1} g^{2-2\alpha_1},
\end{equation}
where we used $g=\gamma_{\rm KN}/\gamma_{\rm b}$. Therefore:
\begin{equation}
\xi \equiv \left[ \frac{3}{4} \left( \frac{m_{\rm e}c^2}{h}\right)^2 \frac{\delta^2}{\nu_{\rm S}\nu_{\rm C}}  \right]^{1-\alpha_1} g^{2-2\alpha_1}.
\end{equation}

\end{document}